\documentclass[12pt]{article}
\usepackage{amsmath,amsfonts,amssymb,dcolumn,lscape,subfigure,alltt,authblk}
\usepackage{graphicx,color}
\usepackage{tabularx}
\usepackage{epstopdf}
\begin{document}

\baselineskip=7.0mm
\setlength\parindent{0pt}
\newcommand{\be} {\begin{equation}}
\newcommand{\ee} {\end{equation}}
\newcommand{\Be} {\begin{eqnarray}}
\newcommand{\Ee} {\end{eqnarray}}
\renewcommand{\thefootnote}{\fnsymbol{footnote}}
\def\a{\alpha}
\def\b{\beta}
\def\g{\gamma}
\def\G{\Gamma}
\def\d{\delta}
\def\D{\Delta}
\def\e{\epsilon}
\def\k{\kappa}
\def\l{\lambda}
\def\L{\Lambda}
\def\t{\tau}
\def\om{\omega}
\def\Om{\Omega}
\def\s{\sigma}
\def\lg{\langle}
\def\rg{\rangle}
\def\et{{\cal E}(t)}
\def\Ge{{\bf G}}

\author[]{Gregor Diezemann}
\affil[]{Department Chemie, Johannes Gutenberg Universit\"at Mainz, Duesbergweg 10-14, 55128 Mainz, Germany}
\title
{Nonlinear response theory for Markov processes IV:\\ The asymmetric double well potential model revisited\\
Revised version
}
\maketitle
\begin{abstract}
\noindent 
The dielectric response of non-interacting dipoles is discussed in the framework of the classical model of stochastic reorientations in an asymmetric double well potential (ADWP).
In the nonlinear regime, this model exhibits some pecularities in the static response.
We find that the saturation behavior of the symmetric double well potential model does not follow the Langevin function
and only in the linear regime the standard results are recovered.
If a finite asymmetry is assumed, the nonlinear susceptibilities are found to change the sign at a number of characteristic temperatures that depend on the magnitude of the asymmetry, as has been observed earlier for the third-order and the fifth-order response.
If the kinetics of the barrier crossing in the ADWP model is described as a two-state model, we can give analytical expressions for the values of the characteristic temperatures.
The results for the response obtained from a (numerical) solution of the Fokker-Planck equation for the Brownian motion in a model ADWP behaves very similar to the two-state model for high barriers.
For small barriers no clearcut time scale separation between the barrier crossing process and the intra-well relaxation exists and the model exhibits a number of time scales.
In this case, the frequency-dependent linear susceptibility at low temperatures is dominated by the fast intra-well transitions and at higher temperatures by the barrier crossing kinetics.
We find that for nonlinear susceptibilities the latter process appears to be more important and the intra-well transitions play only role at the lowest temperatures.
\end{abstract}
%
%
\newpage
\section*{I. Introduction}
Nonlinear (dielectric) spectroscopy is a well established technique to study both, the static and the dynamical behavior of complex systems beyond the linear regime\cite{Rzoska04,Ranko18,Richert:2017}.
In an isotropic system of non-interacting dipoles the static response is given by the Langevin function, which determines the effect of dielectric saturation.
Additionally, the cubic susceptibility of such a 'dipolar gas' is negative, for a discussion of internal field effects see, e.g.\cite{Buchenau:2017}.
Interesting nonlinear effects are to be expected particularly in systems undergoing a phase transition such as, e.g., 
spin glasses, where the nonlinear (magnetic) static susceptibility diverges near the transition temperature\cite{Binder:1986}.
The nonlinear dielectric response has been studied for a variety of different systems, including plastic crystals\cite{Michl:2015},
liquids\cite{Pochec:2019}, or ion conducting materials\cite{Patro:2017}.
In some systems, a change in the sign of the so-called nonlinear dielectric effect (NDE), the static nonlinear susceptibility, has been observed\cite{Rzoska:2018}.

In the field of supercooled liquids and glasses, nonlinear dielectric spectroscopy has been applied to study the details of the heterogeneous slow dynamics in the vicinity of the glass transition.
While nonresonant hole-burning experiments\cite{Schiener:1996,Ralph:2018} have been designed mainly to monitor dynamic heterogeneities, the study of nonlinear susceptibilities aims at identifying spatial correlations in glassy materials, see, e.g., refs.\cite{Richert:2017,Albert:2019} for reviews. 
It was found experimentally that the modulus of the third-order response exhibits a peak-like structure in a frequency range located near the primary relaxation of the supercooled liquid\cite{CrausteThibierge10}.
This so-called hump has been interpreted as originating from glassy correlations as predicted some time before theoretically\cite{Bouchaud05}.
Later, also the fifth-order response has been observed and the results have been discussed in the same framework\cite{Albert:2016}.
The observations are fully in line with the expectations of the random first order theory of the glass transition\cite{Kirkpatrick:1989,Kirkpatrick:2015}.
However, this interpretation of the results apparently is not the only possible one.
Nonlinear susceptiblities computed utilizing concepts of dynamic facilitation theory exhibit hump-like structures in accord with the experimental findings\cite{Royall:2020,Speck:2021}.
Furthermore, a number of models that do not exhibit any spatial aspects of the dynamics have also been shown to exhibit peak-like structures\cite{G75,G81,Kim:2016,Richert:2016,G87,G95}.

A well-known model that has been used in various ways to compute the response to an external field in different physical situations is the model of non-interacting dipoles reorienting in an asymmetric double well potential (ADWP model).
Together with the Debye model of isotropic reorientations it constitutes one of the standard models  of dipole reorientations responsible for dielectric relaxation\cite{Frohlich}. 
The model has also been applied to describe different relaxation processes in glassy systems, 
examples include the (secondary) relaxation in the amorphous state\cite{Gilroy:1981}, the $\b$-relaxation in the supercooled liquid state\cite{Jeppe:2003} or the mechanical relaxation\cite{Buchenau:2001}.
It has furthermore been used for the calculation of nonlinear dielectric response functions, including the modeling of the results of    dielectric holeburning experiments\cite{G39,G46} and the nonlinear response\cite{Buchenau:2017,G75,G87,Ladieu:2012} of glassforming liquids.
In the previous calculations of the third-order and the fifth-order response for the ADWP model, some unusual behavior of the static susceptibilites was observed.
It has been found that the cubic susceptibility changes its sign at a characteristic temperature, while for the fifth-order susceptibility there are two sign-changes at different temperatures\cite{G75,G87}.
 
The intention of the present study was to investigate this behavior in more detail.
In the most part of the paper, we therefore concentrate on the static response functions.
The frequency-dependence of the susceptibilities will be discussed only briefly for a specific ADWP model.
The remainder of the paper is organized as follows.
In the following Section, we briefly outline the computation of nonlinear response functions for stochastic models.
In Section III, we present and discuss the results for the ADWP, both for the two-state model and for the Brownian motion in a bistable potential.
The paper closes with some conclusions.
\section*{II. Nonlinear Response functions}
In this Section, we recapitulate the calculation of the response of a dynamical system with a kinetics described by a master equation (ME)\cite{vkamp81} using the same notation as in refs.\cite{G75,G87}.
Writing $G_{kl}(t,t_0)$ for the conditional probability to find the system in state $k$ at time $t$ provided it was in state $l$ at time $t_0$, the ME has the form
\be\label{ME.t.abh}
\dot G_{kl}(t,t_0)=-\sum_nW_{nk}(t)G_{kl}(t,t_0)+\sum_nW_{kn}(t)G_{nl}(t,t_0)
\ee
where the rates for a transition from state $k$ to state $l$ are given by $W_{lk}(t)$.
The time-dependent populations of the states, $p_k(t)$, obey the same ME and are given by
$p_k(t)=\sum_lG_{kl}(t,t_0)p_l(t_0)$.
The response of the system to an external field $H$ applied at time $t_0$ and measured by a 'moment' $M(t)$, i.e. the polarization, is given by
\be\label{M.expect}
P(t)=\lg M(t)\rg =\biggl\lg\sum_{k}M_k p^{(H)}_k(t)\biggr\rg
= \biggl\lg\sum_{kl}M_k G^{(H)}_{kl}(t,t_0)p_k(t_0)\biggr\rg
\ee
with the vacuum permittivity set to unity.
In eq.(\ref{M.expect}), $p^{(H)}_k(t)$ and $G^{(H)}_{kl}(t,t_0)$ denote the respective quantities in the presence of the field and 
$\lg\cdot\rg$ is an additional average if necessary.
The $M_k$ are the values of the moment in state $k$.
In most relevant situations, one considers systems that are in thermal equilibrium prior to the application of the field, i.e. the initial populations are given by the field-free equilibrium populations $p_k(t_0)=p_k^{\rm eq}$, but other choices are possible as well.
The impact of the coupling to the external field on the transition rates is assumed to originate from a modification of the Boltzmann factors due to the change of the energy of state $k$ by the contribution ($-HM_k$).
A rather general model is obtained using the following expression:
\be\label{Wkl.HX}
W_{kl}^{(H)}(t)=W_{kl}(t)e^{\b H(t)[\g M_k-\mu M_l]}
\ee
Here, $\g$ and $\mu$ can be chosen arbitrarily\cite{G39,CR03,G54} and $\beta=T^{-1}$ with the Boltzmann constant set to unity.
If one chooses $\g=1-\mu$, the model fulfills detailed balance and this will be assumed throughout the present paper.

A perturbation expansion is achieved via the expansion of the transition rates $W_{kl}^{(H)}(t)$ and a concomittant expansion of the propagator, i.e. the matrix ${\bf G}^{(H)}(t,t_0)$, in terms of the corresponding 'field-free' propagator ${\bf G}(t,t_0)$ using the decomposition of the matrix of transition rates, ${\cal W}^{(H)}(t)={\cal W}(t)+{\cal V}(t)$, with 
${\cal V}(t)=\sum_{n=1}^\infty{\cal V}^{(n)}(t)$.
The definition of the $n$th order perturbation term ${\cal V}^{(n)}(t)$ results from the Taylor series of the factor 
$e^{\b H[\g M_k-\mu M_l]}$ in eq.(\ref{Wkl.HX}), cf.\cite{G75,G81}.
Starting from the general perturbation expansion, the approximations to the propagator are found from
\be\label{Gn.allg}
\Ge^{(n)}(t,t_0)=\sum_{m=0}^{n-1}\int_{t_0}^t\!dt'\Ge(t,t'){\cal V}^{(n-m)}(t')\Ge^{(m)}(t',t_0)
\ee
cf. refs.\cite{G87,G92}.
The terms of third- and fifth-order are explicitly given in ref.\cite{G87}.
(We mention that there the term involving only ${\cal V}^{(1)}$ is erroneously missing in the expression for the third-order Green's function.)

The polarization is then given as a sum of nonlinear susceptibilities
\[
P(t)=\!\!\!\sum_{n\,\rm uneven}\!\!\!H^n\chi_n(\om)
\]

The frequency-dependent nonlinear susceptibilities of third- and fifth-order for a number of stochastic models have been calculated this way\cite{G75,G87,G95,G92}.

In the particular case of a Fokker-Planck equation (FPE) the perturbation expansion for the propagator simplifies 
considerably and it reduces to the one already derived a long time ago by Morita\cite{Morita:1986,G92}.
To show this, one starts from the ME representing a so-called one-step process, i.e. a process with transitions only among nearest neighbor states characterized by discrete values $q_k$ of a coordinate $q$, and relates this to the FPE for the Brownian motion in a potential by choosing for the transition rates\cite{vkamp81,Agmon:1983}: 
\be\label{W.diff}
\bar W_{k(k\pm1)}=\bar{D}e^{-(\b/2)(V(q_k)-V(q_{k\pm1}))}
\simeq \bar{D}\left[1-(\b/2)(V(q_k)-V(q_{k\pm1}))\right]
\ee
Here, $\bar{D}=D/\D_q^2$ with $D$ denoting the diffusion coefficient and $\D_q=(q_{k+1}-q_k)$ is the spacing of the discrete representation of the coordinate $q$ in the limit of vanishing $\D_q$.
If the potential includes the term $(-M_k\cdot H)$, it is obvious that only the linear coupling to the field is relevant in the continuum limit.
This is easily obtained from eq.(\ref{Wkl.HX}) for $\g=\mu=1/2$ if one substitutes $W_{kl}(t)$ by $\bar W_{k(k\pm1)}\d_{k,(l\pm1)}$ according to eq.(\ref{W.diff}).
This way, one shows that only the terms ${\cal V}^{(1)}$ are finite and all other terms in the perturbation expansion 
(\ref{Gn.allg}) vanish.
\section*{III. Nonlinear response in the ADWP model}
We will discuss the results for nonlinear response functions obtained for the ADWP model in two different ways.
When one starts from a two-state model, the nonlinear response functions can be computed analytically see, refs.\cite{G75,G87}.
If, on the other hand, the Brownian motion in a model bistable potential is considered, one has to rely on numerical solutions of the FPE and use the results in the perturbation expansion of the propagator.
Apart from this difference in the technical details of the calculations, using the two-state model for the kinetics in an ADWP is meaningful only if there is a time scale separation between the intra-well kinetics and the inter-well transitions.
With the two-state model only the latter are treated properly while the solution of the FPE for the Brownian motion in a bistable potential allows to discuss both processes.
\subsubsection*{A. The two-state ADWP model}
\begin{figure}[h!]
\centering
\includegraphics[width=7.0cm]{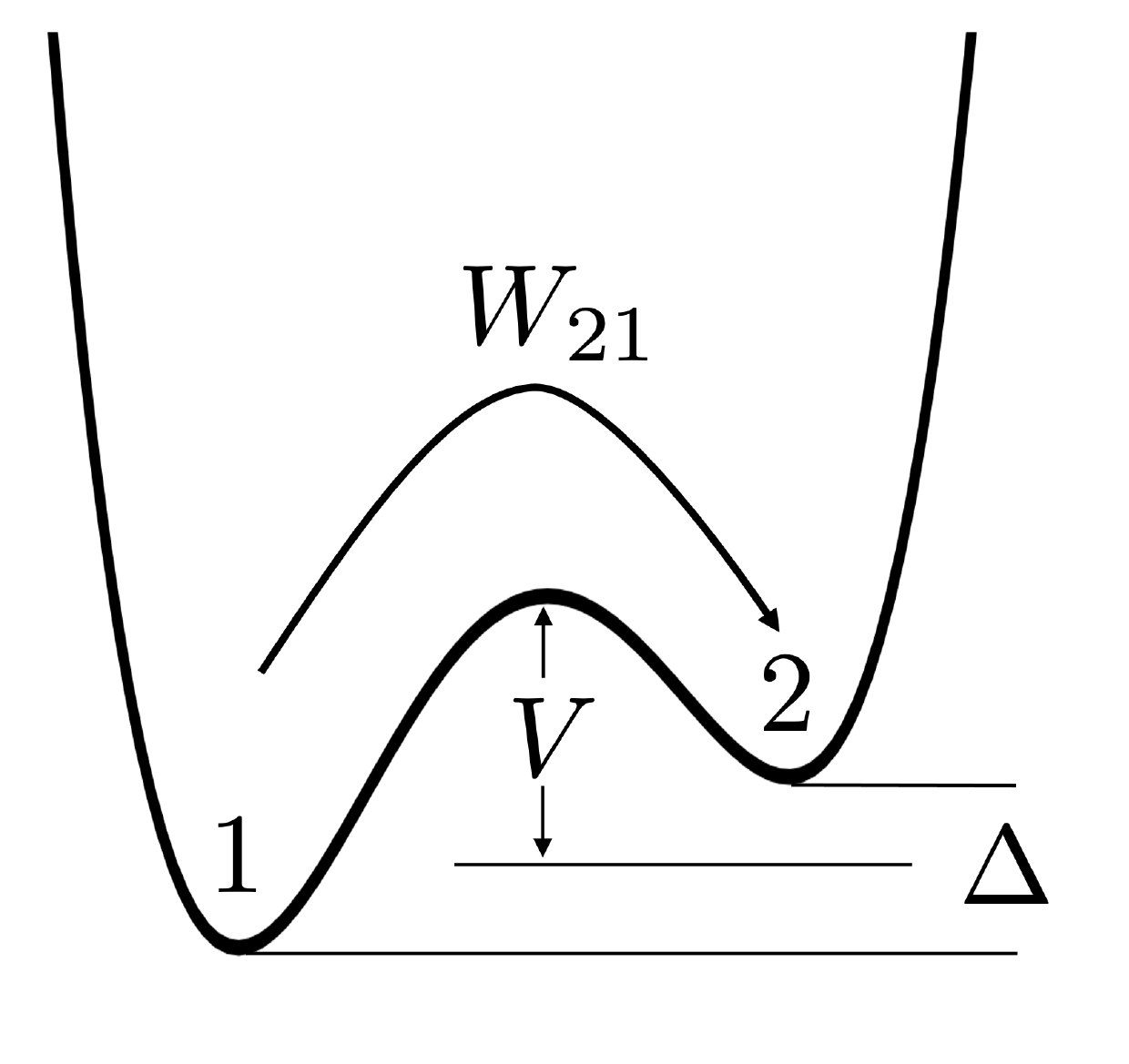}
\vspace{-0.5cm}
\caption{Sketch of an ADWP, indicating the transition rate between the two states. $V$ is the barrier and $\D$ denotes the asymmetry. For the plot, I used a potential of the form $V(q)=(1/4)q^4-(1/2)q^2+(1/12)q^3$.}
\label{Fig1}
\end{figure}

The two-state model for dipole reorientation in an ADWP can be summarized in the following way, cf. refs.\cite{G75,G87}.
The minima of the potential are associated with two dipole orientations characterized by polar angles $\theta_1=\theta$ and 
$\theta_2=\theta+\pi$ cf. Fig.\ref{Fig1}.
For the coupling to the field $(-M\cos{\!(\theta_k)}\cdot H$) the values of the moments associated with each well,
$M_k=M\cos{\!(\theta_k)}$, i.e. $M_1=M\cos{\!(\theta)}$ and $M_2=-M\cos{\!(\theta)}$, are relevant. Here, $M$ denotes the static molecular dipole moment.
If one treats liquid systems, usually isotropic distributions of the moments are assumed and the angular average yields 
$\lg\cos^n{\!(\theta)}\rg=(n+1)^{-1}$ for $n$ even and the average vanishes otherwise.

The transitions between the wells take place with rates $W_{21}=W_{2\leftarrow1}=We^{-\b\D/2}$ and
$W_{12}=We^{+\b\D/2}$, where $\D$ denotes the asymmetry and the bare rate for $\D=0$ is given by $W=W_0e^{-\b V}$.
For this model, the ME, eq.(\ref{ME.t.abh}), can be easily solved analytically. In the field-free situation, one has
$G_{kl}(t)=p_k^{\rm eq}\left(1-e^{-R\cdot t}\right)+\d_{kl}e^{-R\cdot t}$
with the relaxation rate
\be\label{R.H0}
R=W_{12}+W_{21}=2W\cosh{\!(\b\D/2)}
\ee
and the equilibrium populations $p_k^{\rm eq}=W_{kl}/R$.

One can give an analytical solution of the ME also in the presence of an external field.
Since $M_2=-M_1$, one has $W_{kl}^{(H)}(t)=W_{kl}e^{\b H(t) M_k}=W_{kl}e^{-\b H(t) M_l}$ for $k,l=1,2$, 
independent of the parameters $\g$ and $\mu=(1-\g)$, cf. eq.(\ref{Wkl.HX}).
The relaxation rate in this case is given by
\be\label{RH.t}
R_H(t)=2W\cosh{(\b\et)}\quad\mbox{with}\quad\et=\D/2+H(t)\!\cdot\! M\!\cdot\! \cos{(\theta)}.
\ee
From the solution of the ME, one can compute the polarization, eq.(\ref{M.expect}), with the result:
\be\label{Mav.t.gen}
P(t) = M\left\lg\cos{(\theta)}
e^{-\G_H(t,t_0)}\left\{\d + \int_{t_0}^t\!dt'e^{\G_H(t',t_0)} \tanh{(\b{\cal E}(t'))}\right\}
\right\rg
\ee
Here, the angular bracket includes an angular average and we assumed that the field is switched on at time $t_0$ and the system was in thermal equilibrium before.
Additionally, we defined 
\be\label{delta.def}
\d=\tanh{\!(\b\D/2)}
\ee
and $\G_H(t,t_0)=\int_{t_0}^t\!dt' R_H(t')$.

Eq.(\ref{Mav.t.gen}) can be used to compute the response in any desired order and the results coincide with those obtained from the general perturbation expansion discussed in refs.\cite{G75,G87}.
The linear response is given by the well known Debye-like expression $\chi_1(\om)=\D\chi_1[1-i\om\t]^{-1}$, with the static susceptibility $\D\chi_1=\b(M^2/3)\left(1-\d^2\right)$ and $\t=R^{-1}$.
Additionally, the nonlinear response functions of third-order and fifth-order have been discussed.
In particular, it has been found that there is a peak in the moduli observable in the vicinity of some characteristic temperatures which are defined by a vanishing zero frequency limit of the susceptibilites. 
For $\chi_3(\om)$, there is one characteristic temperature and for $\chi_5(\om)$ it are two:
\Be\label{T.3.5}
T_3\simeq0.759\D\quad\mbox{and}\quad
T_{5;1}\simeq0.318\D \,;\, T_{5;2}\simeq1.187\D
\Ee

In order to discuss the zero-frequency nonlinear susceptibilities and the saturation properties of the model in further detail,
it is sufficient to consider the response to a d.c. field $H=H\theta(t-t_0)$.
The static response is determined by the long-time limit of eq.(\ref{Mav.t.gen}) and one explicitly finds:
\be\label{Chi.stat}
P_{\rm ADWP}=M\bigl\lg\cos{(\theta)}\tanh{\left[\b(\D/2+H\!\cdot\! M\cos{(\theta)})\right]}\bigr\rg
\ee
We now expand this expression in powers of $H$, $P_{\rm ADWP}=\sum_{n}\!P_n=\sum_{n}\!H^n\D\chi_n$, and find that in each non-vanishing order ${\cal O}(H^n)$ there exist $(n-1)/2$ characteristic temperatures $T_{n,\a}$, depending on the value of the asymmetry $\D$, for which the static susceptibility vanishes (and changes sign).
For vanishing asymmetry the $T_{n,\a}$ vanish, cf. eq.(\ref{Tna.def}).
The actual calculation up to $n=9$ is outlined in Appendix A.
There, also the expressions for the temperatures $T_{7,\a}$ and $T_{9,\a}$ are given (eq.(\ref{T.7.9})).
In this context, it is to be noted that the results for $\D\chi_n$ coincide with the sum of all the zero-frequency limits,
e.g. $\D\chi_3=(3/4)\chi_3^{(1)}(0)+(1/4)\chi_3^{(3)}(0)$, cf. refs.\cite{G75,G87} and Appendix B.
In ref.\cite{G87} only the component $\chi_5^{(5)}(\om)$ was calculated and its zero-frequency limit is only $1/16$ of $\D\chi_5$ computed here, cf.\cite{Albert:2019,Albert:2016}.

\begin{figure}[h!]
\centering
\includegraphics[width=8.0cm]{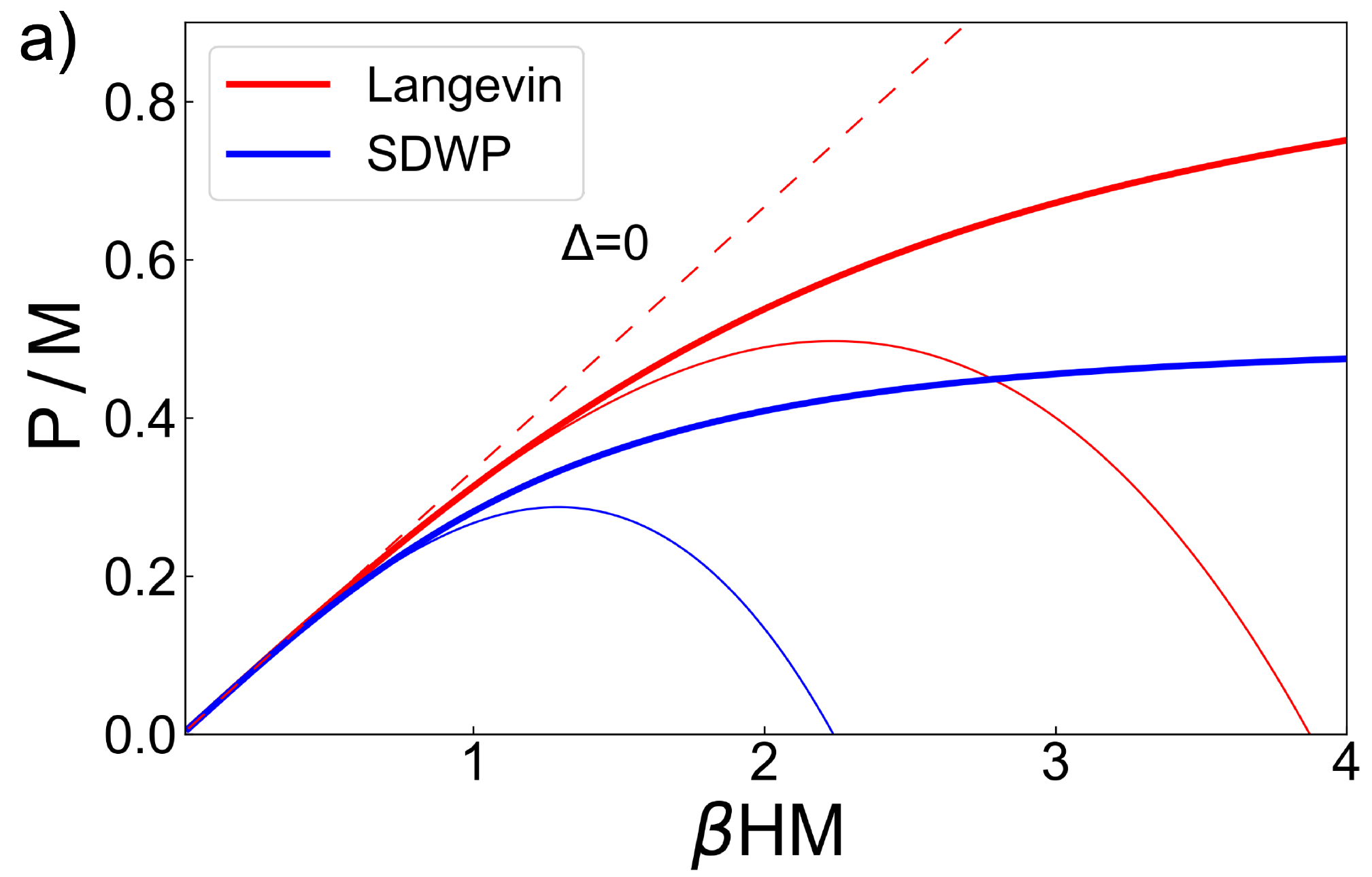}
\includegraphics[width=7.9cm]{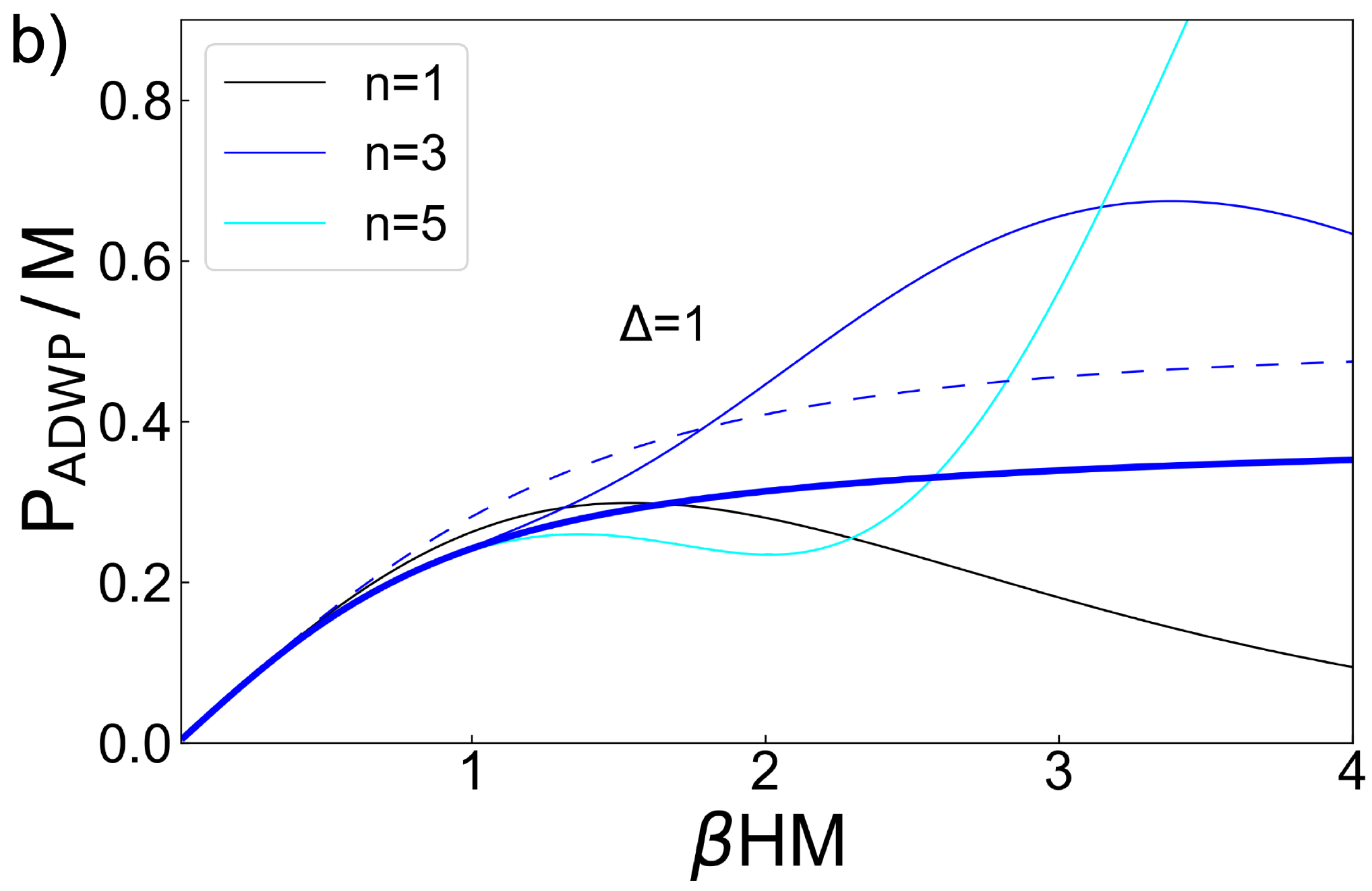}
\vspace{-0.5cm}
\caption{{\bf a)} Static polarization for the SDWP model ($\D=0$) and the behavior according to the Langevin function. 
The dashed line is the linear approximation and the thin lines represent the Taylor expansions up to order $(\b H M)^3$, i.e.
for the Langevin function ($x/3-x^3/45$) and for the SDWP ($x/3-x^3/15$).
{\bf b)} Static polarization for $\D=1$ as a function of $(\b H M)$.
Thin lines represent the approximations up to the given order in the field strength, using the expressions given in Appendix A.
The thin dashed blue line is $P_{\rm SDWP}$, cf. a).
}
\label{Figure2}
\end{figure}

We start the discussion of eq.(\ref{Chi.stat}) considering a symmetric double well potential (SDWP) model, i.e. $\D=0$.
For a three-dimensional system, the equilibrium polarization according to elementary statistical mechanics is given by the Langevin function\cite{Frohlich,Speck:2021}:
\[
P_{\rm eq}/M=L(x)=\coth{(x)}-{1\over x}={1\over3}x-{1\over45}x^3+{2\over945}x^5-{1\over4725}x^7+\cdots
\]
where $x=\b H M$.
For the SDWP model we have from eq.(\ref{Chi.stat}) and the expressions given in Appendix A:
\[
P_{\rm SDWP}/M={1\over3}x-{1\over15}x^3+{2\over105}x^5-{17\over2835}x^7+\cdots
\]
In Fig.\ref{Figure2}a) we plot $P$ according to the Langevin function and for the SDWP model as a function of the ratio of dipolar energy and thermal energy, 
$\b H M$. 
The different saturation behavior is evident from that figure.
In addition, the behavior up to third order in $H$ is shown for both models (thin lines).
Only in the linear regime all expressions coincide.
(One might argue that this is to be expected as the model is essentially a one-dimensional model but to the best of my knowledge it has not been recognized sofar.)
Experimental results on isotropic systems are usually interpreted in terms of the Langevin function corrected for interactions of the dipoles.
A recent example is provided by a study of the nonlinear dielectric response of propylene glycol, where the so-called Piekara factor characterizing the third-order susceptibility could be determined\cite{Weinstein07}.

The situation changes if the asymmetry is finite, cf. Fig.\ref{Figure2}b), where we plot $P_{\rm ADWP}$ for an asymmetry of $\D=1$.
It is evident that for this example the total polarization behaves qualitatively similar to what is observed in case of the SDWP model and the Langevin function. 
However, the various contributions show a rather different dependence on $(\b H M)$.
The linear response exhibits a maximum and asymptotically decays to zero, cf. the factor $(1-\d^2)$ in the expression for $\D\chi_1$.
Also in third-order a maximum is observed and $\D\chi_3$ asymptotically vanishes.
This anomalous behavior indicates a positive contribution in ${\cal O}(x^3)$ instead of a negative one.
In addition, the changes in slope in fifth-order can be observed, cf. eq.(\ref{Dchi.k}). 

\begin{figure}[h!]
\centering
\includegraphics[width=7.9cm]{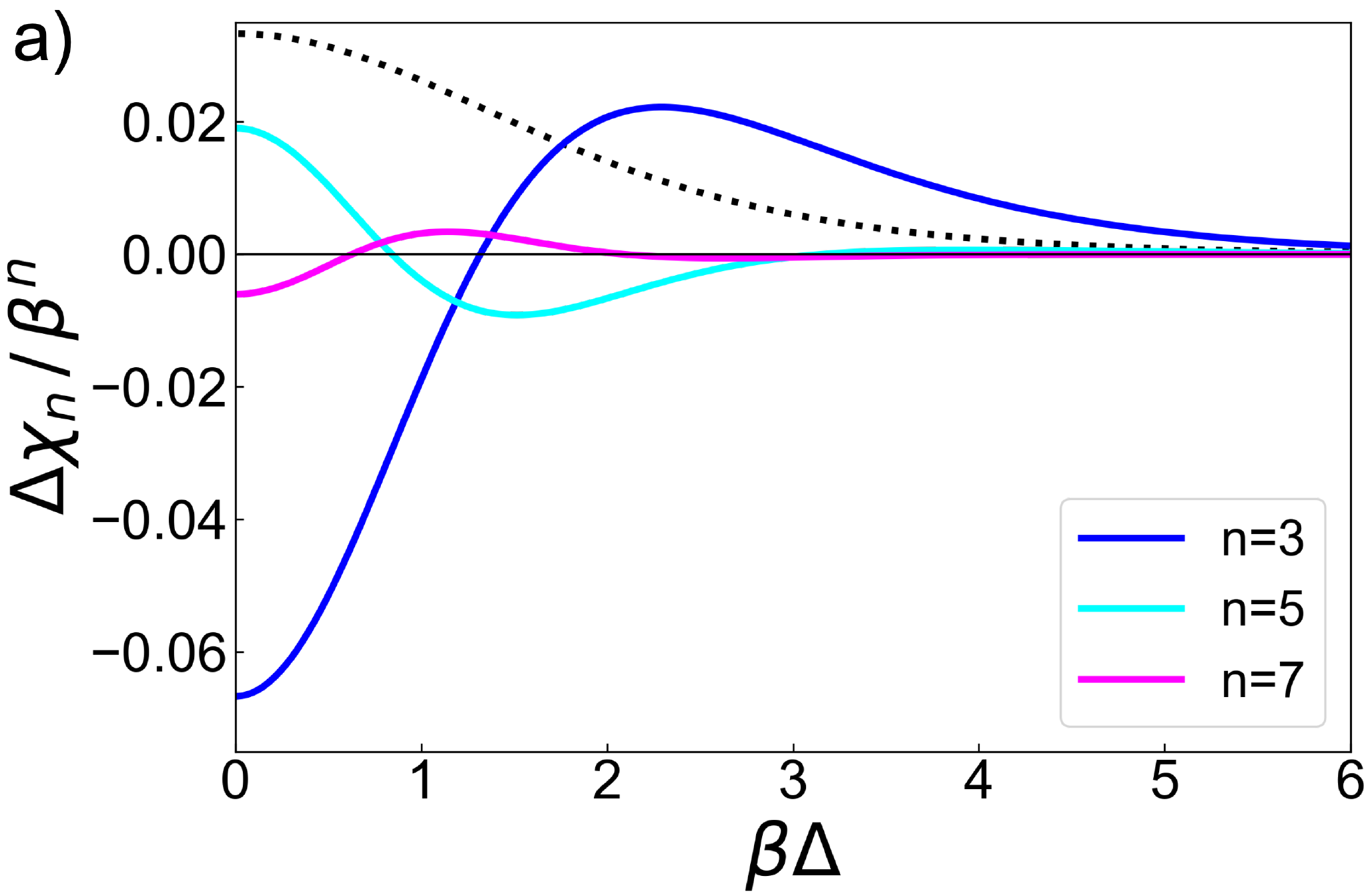}
\includegraphics[width=8.0cm]{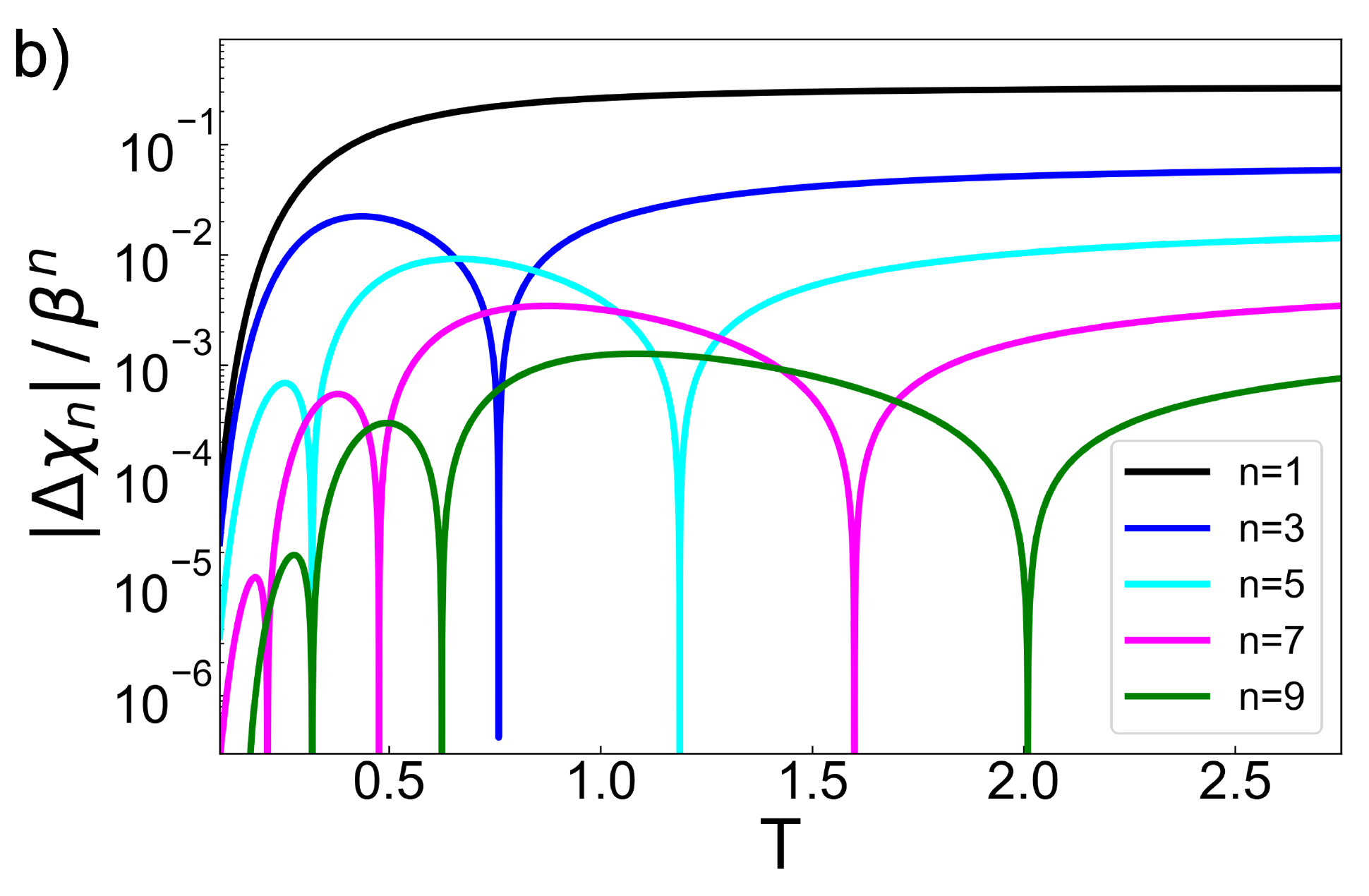}
\vspace{-0.5cm}
\caption{{\bf a)} Static susceptibilities $\D\chi_n\,/\,\b^n$ ($n=3,5,7$) as a function of $\b\D$.
The dotted black line is $\D\chi_1/\b$ divided by a factor of 10.
{\bf b)} Absolute values $|\D\chi_n|\,/\,\b^n$ ($n=1,3,5,7,9$) as a function of temperature for $\D=1$. 
Note that not all $T_{n,\a}$ are covered in the temperature range shown.
}
\label{Figure3}
\end{figure}

The change in sign of the various contributions $\D\chi_n$ is presented in Fig.\ref{Figure3}a) as a function of the product of inverse temperature and asymmetry.
As noted already in the Introduction in some systems a positive NDE (or Piekara factor) has been observed. 
Examples are polar liquids\cite{Piekara:1959,Drozd-Rzoska:2008}, mixtures\cite{Zboinski:1979}, and also systems in the vicinity of a liquid-solid phase transition\cite{Pochec:2019}.
The third-order susceptibility $\D\chi_3$ can be related to the correlation coefficient for saturation, 
$R_S=-(45/M^4)\times(\D\chi_3/\b^3)$ if local field effects are neglected\cite{Rzoska:2018}.
The blue curve in Fig.\ref{Figure3}a) looks similar to what is expected for the so-called chemical effect in liquid mixtures giving rise to a positive NDE\cite{Ranko18,Piekara:1959,Zboinski:1979}.
It would be interesting to see if higher-order susceptiblities of, e.g. liquid mixtures, exhibit more than one sign-change such as $\D\chi_5$ (cyan line).

The general behavior of the higher-order susceptibilities and the sign-changes at the characteristic temperatures $T_{n,\a}$ becomes most obvious if one plots the absolute values $|\D\chi_n|$ on a logarithmic scale as a function of temperature, cf. Fig.\ref{Figure3}b).

\begin{figure}[h!]
\centering
\includegraphics[width=8.0cm]{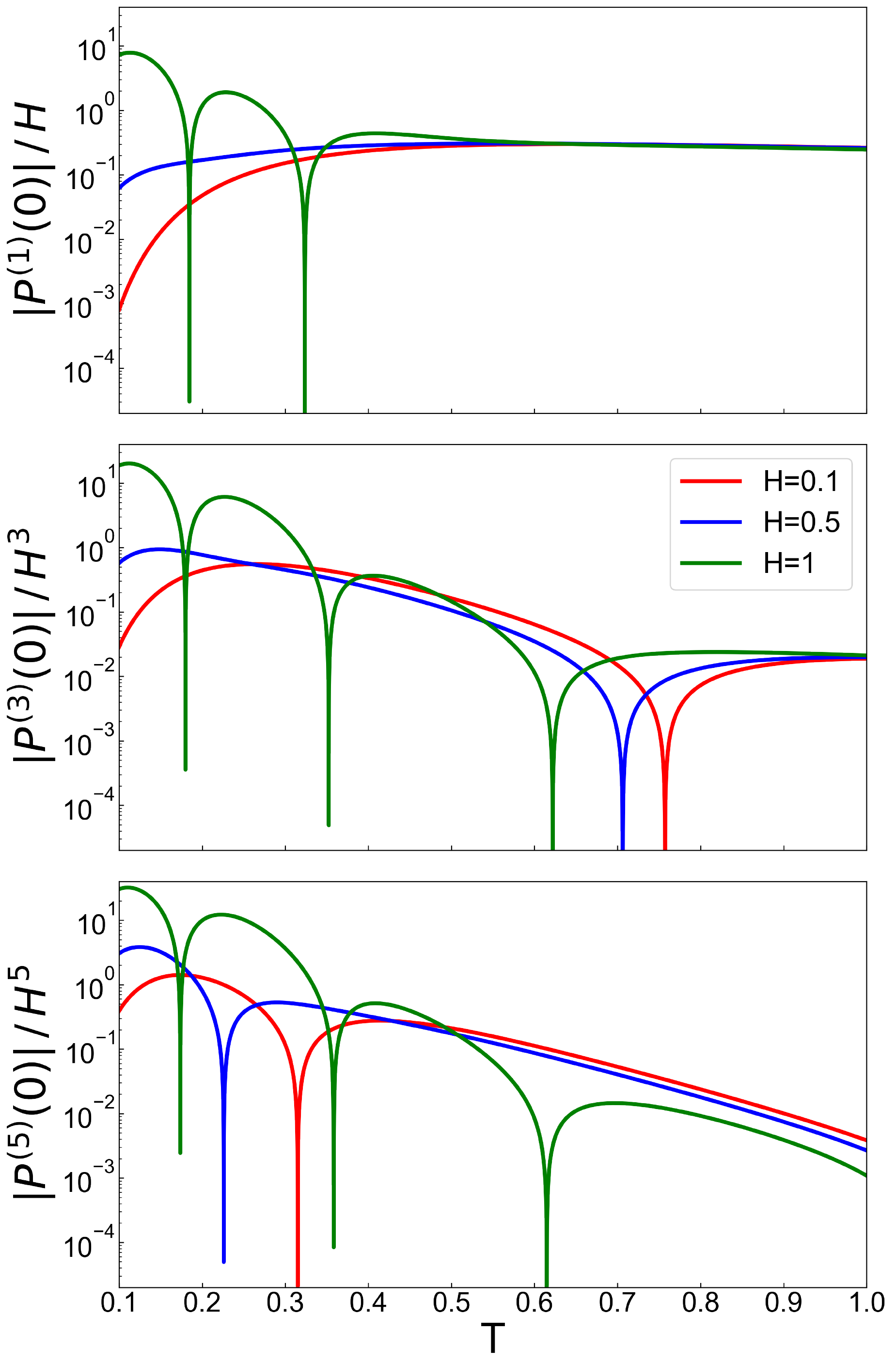}
\vspace{-0.25cm}
\caption{$|P^{(k)}(0)|/H^k$, $k=1,3,5$, as a function of temperature ($\D=1$). 
For $H=0.1$ (red lines), the characteristic temperatures coincide with $T_{n,\a}$. 
}
\label{Figure4}
\end{figure}

It has been noticed in the previous work\cite{G75,G87,Ladieu:2012,Buchenau:2017} that the occurence of a hump in the frequency-dependent nonlinear susceptibilities of a two-state ADWP model is closely related to the vanishing static susceptibility at the respective characteristic temperatures.
The fact that $\chi_n^{(k)}(0)$ vanishes at $T_{n,\a}$ implies that in the vicinity of this temperature the impact of 
$\chi_{n+2}^{(k)}(0)$ cannot necessarily be neglected.
This is exemplified in Fig.\ref{Figure4}, where $|P^{(k)}(0)|$ according to eq.(\ref{Chi.comp}) is shown as a function of temperature for different field strengths.
It is evident, that the impact of the higher-order susceptibilities changes the value of the characteristic temperature and for very strong fields also the number of zeros (sign-changes).
Therefore, a comparison to experimental data exhibiting a hump in the frequency-dependent susceptibilities appears at least to be challenging.

An important aspect of the findings presented in this Section is the following.
If the ADWP model is used to describe the reorientational dynamics of non-interacting dipoles, the equilibrium properties should be determined by the Langevin function according to equilibrium statistical mechanics. 
In this case, the third-order nonlinear response is negative and one does not expect a change of sign as a function of temperature or field strength.
A positive third-order response or anomalous nonlinear dielectric effect can be observed in a number of systems.
These systems, however, have in common that their saturation behavior cannot be described by a model of non-interacting dipoles.
One prominent example is provided by strong interactions among the dipoles in systems that underly special physical conditions like, e.g., the neighborhood of a phase transition or similar phenomena, but other sources for an anomalous behavior have also been discussed, see, e.g., ref.\cite{Richert:2017}.
In addition, the clearcut extraction of the various frequency components of the polarization might be difficult near characteristic temperatures.
\subsubsection*{B. The Fokker-Planck equation for the ADWP model}
The two-state ADWP model can be derived from the FPE for the Brownian motion in a bistable potential, see e.g.\cite{Schulten:1981}.
In order to see whether or not the occurence of the sign-changes of the higher-order susceptibilities is a generic feature of the diffusive barrier crossing kinetics in bistable systems, we consider the Brownian motion of a dipole with coordinate $q$, i.e. $M(q)$, in a model ADWP $V_{ADWP}(q)$ in the presence of an external field $H(t)$. 
The overall potential is given by
\be\label{V.qt}
V(q,t)=V_{ADWP}(q)-M(q)\cdot H(t)
\ee
The stochastic motion of $q(t)$ can either be described using the corresponding Langevin equation or equivalently the FPE:
\be\label{FP.eq}
{\dot G}(q,t|q_0)
=D\left[\partial_q e^{-\b V(q,t)}\partial_qe^{\b V(q,t)}\right]G(q,t|q_0)
\ee
Here, the diffusion coefficient is related to the damping constant, $D=\g T$.
As has been explained above, the response theory for a stochastic dynamics described by a FPE is very similar to the one for a ME
but the response functions consist of less terms.
In the actual calculations, we employed an ADWP of the following form:
\be\label{V.ADWP}
V_{ADWP}(q)=V_0\left({k_4\over4}q^4-{k_2\over2}q^2+{k_3\over3}q^3\right)
\ee
Some specific properties and the relation of the potential parameters to the Kramers rate\cite{vkamp81} are given in Appendix C.
For all model calculations we set $k_2=k_4=1$.

\begin{figure}[h!]
\vspace{-0.25cm}
\centering
\includegraphics[width=8.0cm]{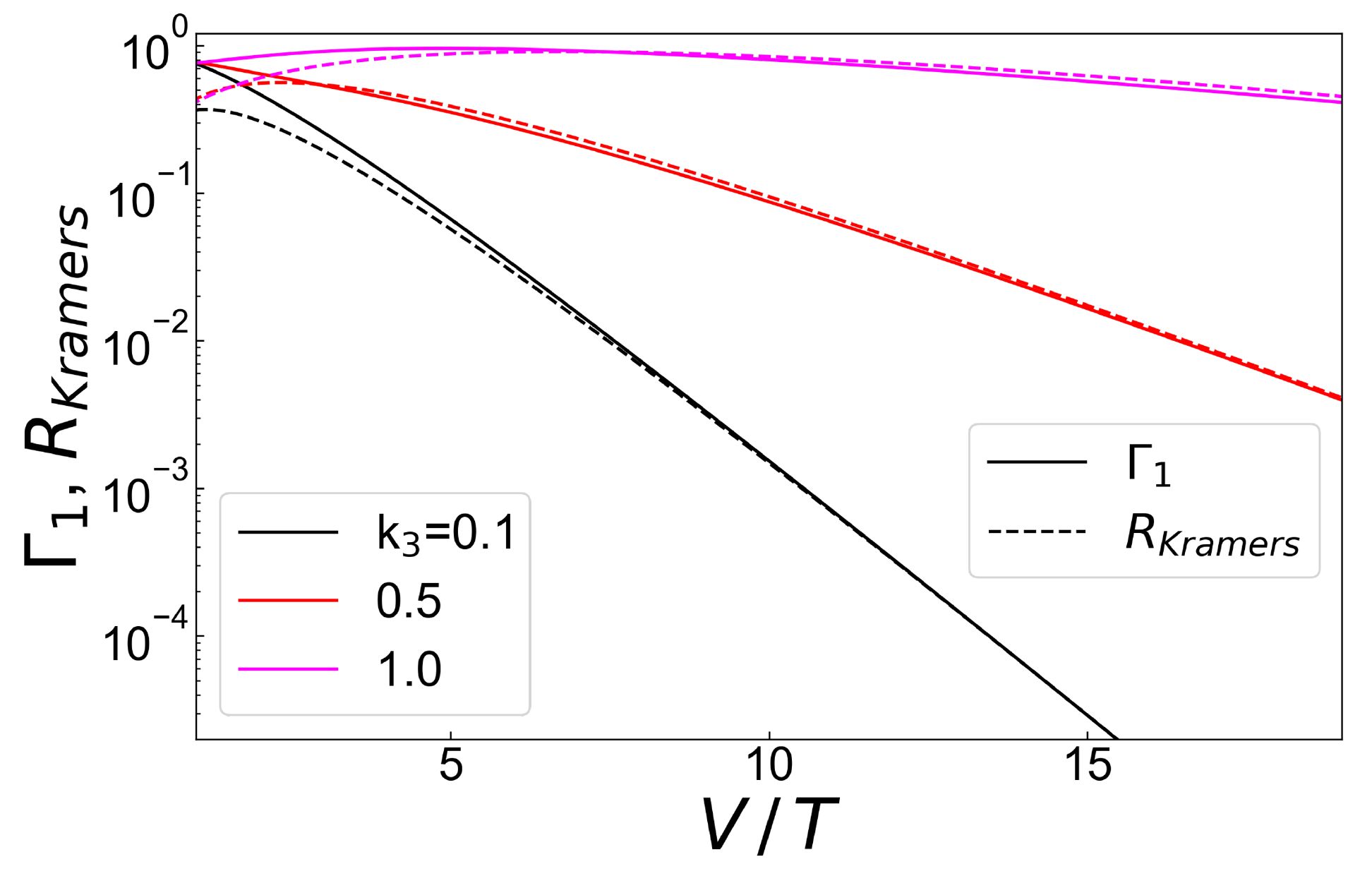}
\vspace{-0.5cm}
\caption{Comparison of the rate $\G_1=-\l_1$ and the rate in the Kramers approximation for the mean first passage time,
$R_{Kramers}$ for different values of $k_3$ determining the asymmetry as a function of the barrier height $V$.}
\label{Figure5}
\end{figure}

It is well known that in case of high barriers the largest non-zero eigenvalue of the FP operator, $\l_1$, is the negative of the rate for the inter-well transitions, $\G_1=-\l_1$.
In Fig.\ref{Figure5}, we show these rates for some values of the parameter $k_3$ as a function of the barrier height $V$ in units of temperature.
In addition to these rates, obtained from a numerical solution of the FPE, the rates in the Kramers approximation
$R_{\rm Kramers}$ are shown for the same parameters, cf. Appendix C.
It is obvious, that these rates coincide with the exact ones for high barriers and not too large $k_3$.
For barrier heights smaller than roughly $10$, there are some systematic discrepancies.
This is also the regime, in which there is no clearcut time scale separation between intra-well and inter-well transitions.

For the computation of the dielectric response, we use a linear relationship for the dependence of $M(q)$ on the reaction coordinate:
\be\label{M.lin.q}
M(q)=-M_0\cdot q\quad\mbox{with}\quad M_0=M\cdot\cos{(\theta)}
\ee
We define the dependence to be the negative of the coordinate in order to assure that the cumulated value of the moment in well '1' is positive in accordance with the definitions used in the two-state model.
The particular choice (\ref{M.lin.q}) has to be viewed as an additional model assumption.

{\bf 1. Statics}\\
In order to assure that the occurence of the characteristic temperatures at which the static susceptibilities change their sign
is not an artefact of the two-state model, we have computed $\D\chi_n$ for the quartic ADWP given in eq.(\ref{V.ADWP}).
Some details of the calculations are outlined in Appendix C.

In Fig.\ref{Figure6}a), we present the results for $|\D\chi_3|$ and $|\D\chi_5|$ as the full lines. 
\begin{figure}[h!]
\begin{minipage}{0.475\textwidth}
\centering
\vspace{-7.0cm}
\subfigure[$|\D\chi_n|$ for $\b V_0=50$, $k_3=0.0299$ ($\D=1$).]
{\includegraphics[width=\textwidth]{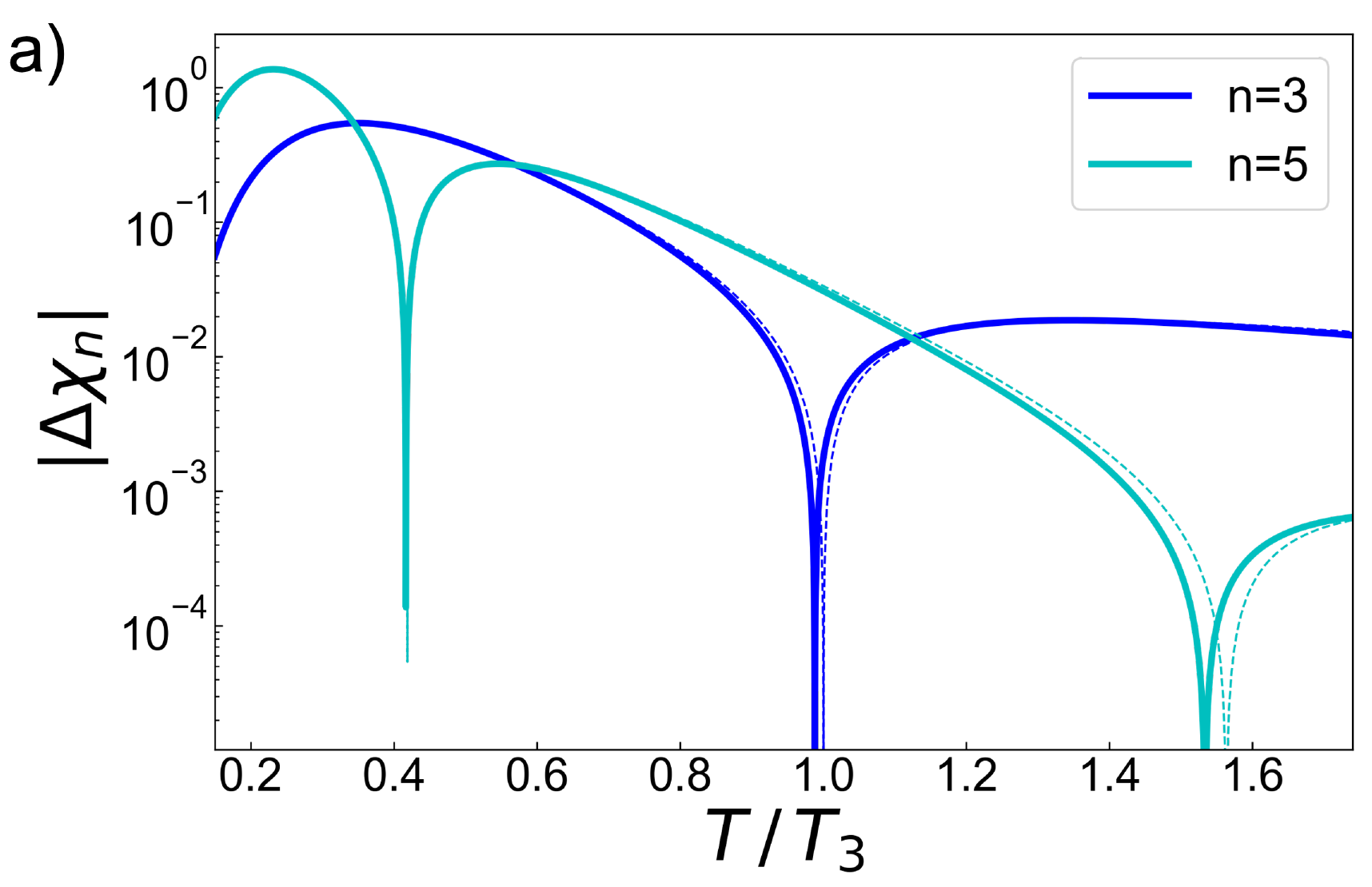}}
\end{minipage}
\begin{minipage}{0.475\textwidth}
\subfigure[$|\D\chi_n|$ for $\b V=10$ and various $k_3$.
According to eq.(\ref{V.1.2}), one has for $k_3=0.1$: $\b V_0=39.6$, $\D=2.65$,
for $k_3=0.5$: $\b V_0=31.7$, $\D=11.57$, and for $k_3=1.0$: $\b V_0=18.5$, $\D=17.24$.]
{\includegraphics[width=\textwidth]{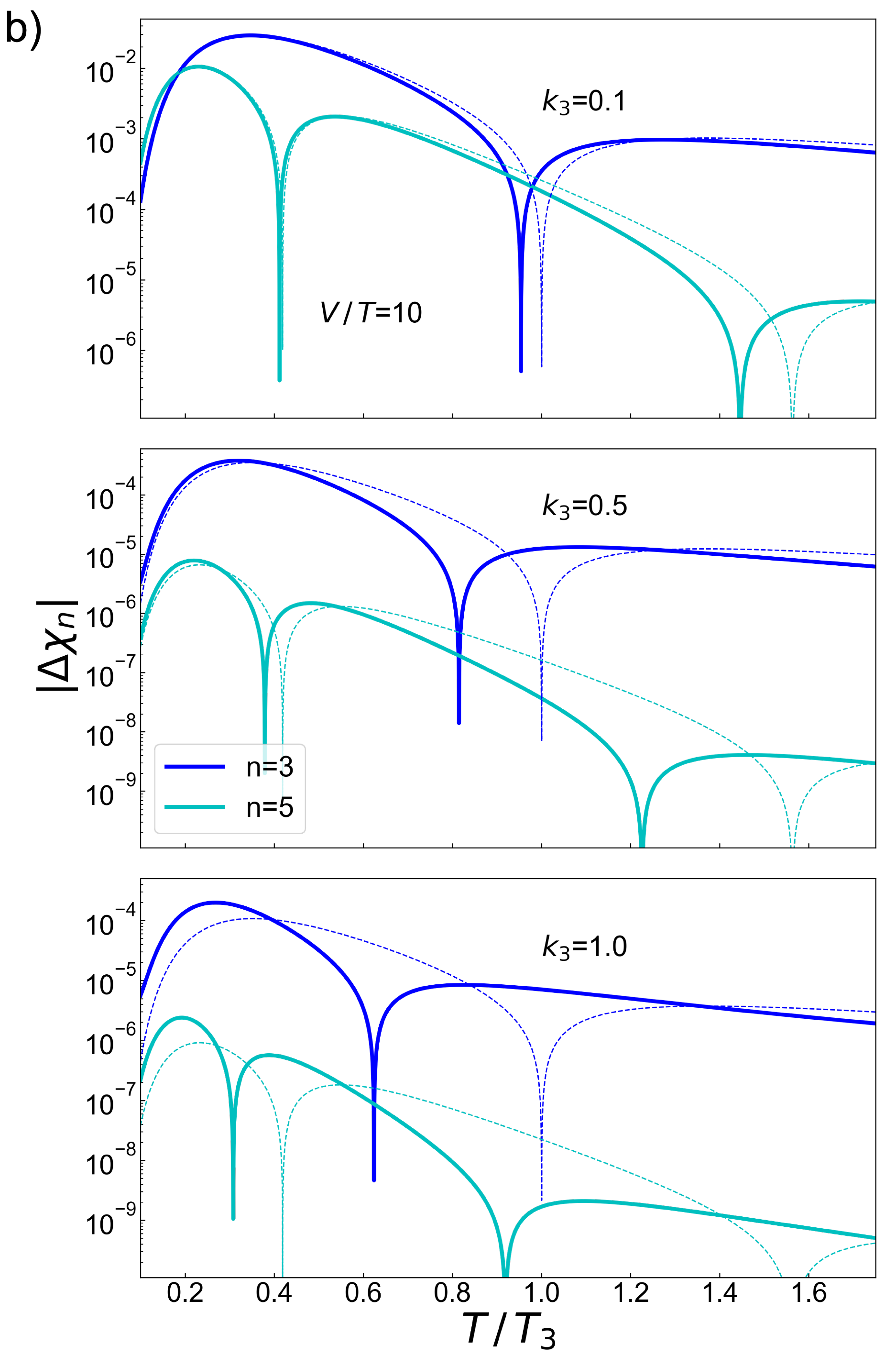}}\\
\end{minipage}
\begin{minipage}{0.475\textwidth}
\vspace{-7.5cm}
\subfigure[Temperature of vanishing $\D\chi_3$, $T(\D\chi_3=0)$, scaled to $T_3$ (eq.(\ref{T.3.5})) as a function of the potential height for various values of $k_3$.]
{\includegraphics[width=\textwidth]{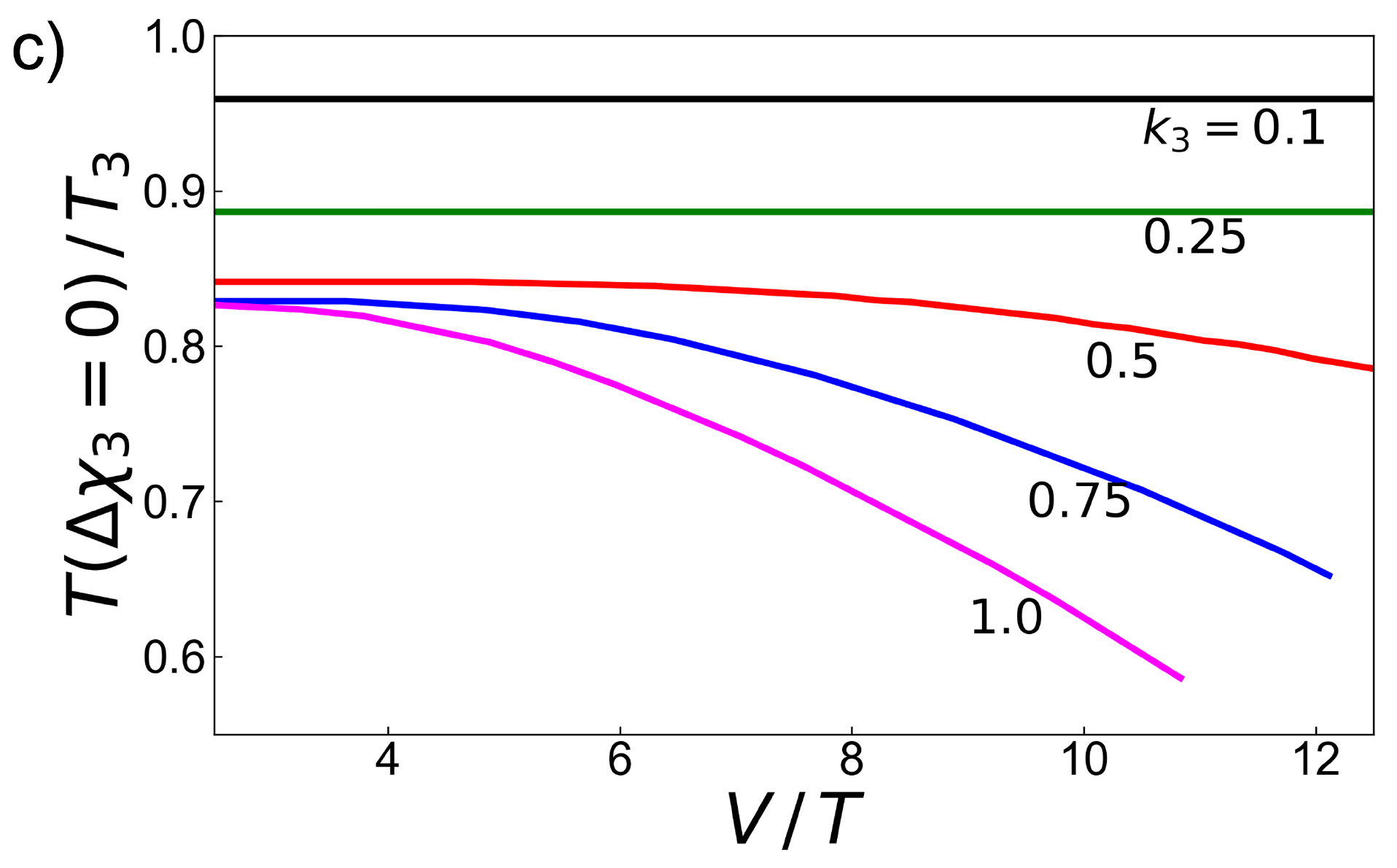}}
\end{minipage}
\caption{Static susceptibilities $|\D\chi_n|$ according to eq.(\ref{DChi.n.FPE}) as a function of temperature scaled to $T_3$, 
cf. eq.(\ref{T.3.5}), and characteristic temperatures for $k_2=k_4=1$ and various values of $k_3$ as indicated.
Dashed lines represent the results obtained for the two-state model.}
\label{Figure6}
\end{figure}
These calculations show that the existence of the zeros in the $\D\chi_n$ are not an artefact of the approximations inherent in the two-state model, but also occur in the original model of stochastic dynamics in an ADWP.

Next, we consider the dependence of the susceptibilities on the potential barrier including the regime of small barriers.
In Fig.\ref{Figure6}b), we show results for $|\D\chi_n|$ for $\b V=10$ and varying $k_3$.
It is evident that with increasing asymmetry (larger value of $k_3$) the characteristic temperatures change towards smaller values.
This is at variance with the results for the two-state model, eq.(\ref{Tna.def}), where $T_{n,\a}$ increases with increasing $\D$.
However, the overall trends in the behavior of $|\D\chi_n|$ still are reasonably met by the two-state model. 
In particular, the number of sign-changes is not altered.
Without showing the results here, we mention that this behavior is also found for $\D\chi_n$ with larger $n$. 

\begin{figure}[h!]
\centering
\includegraphics[width=8.0cm]{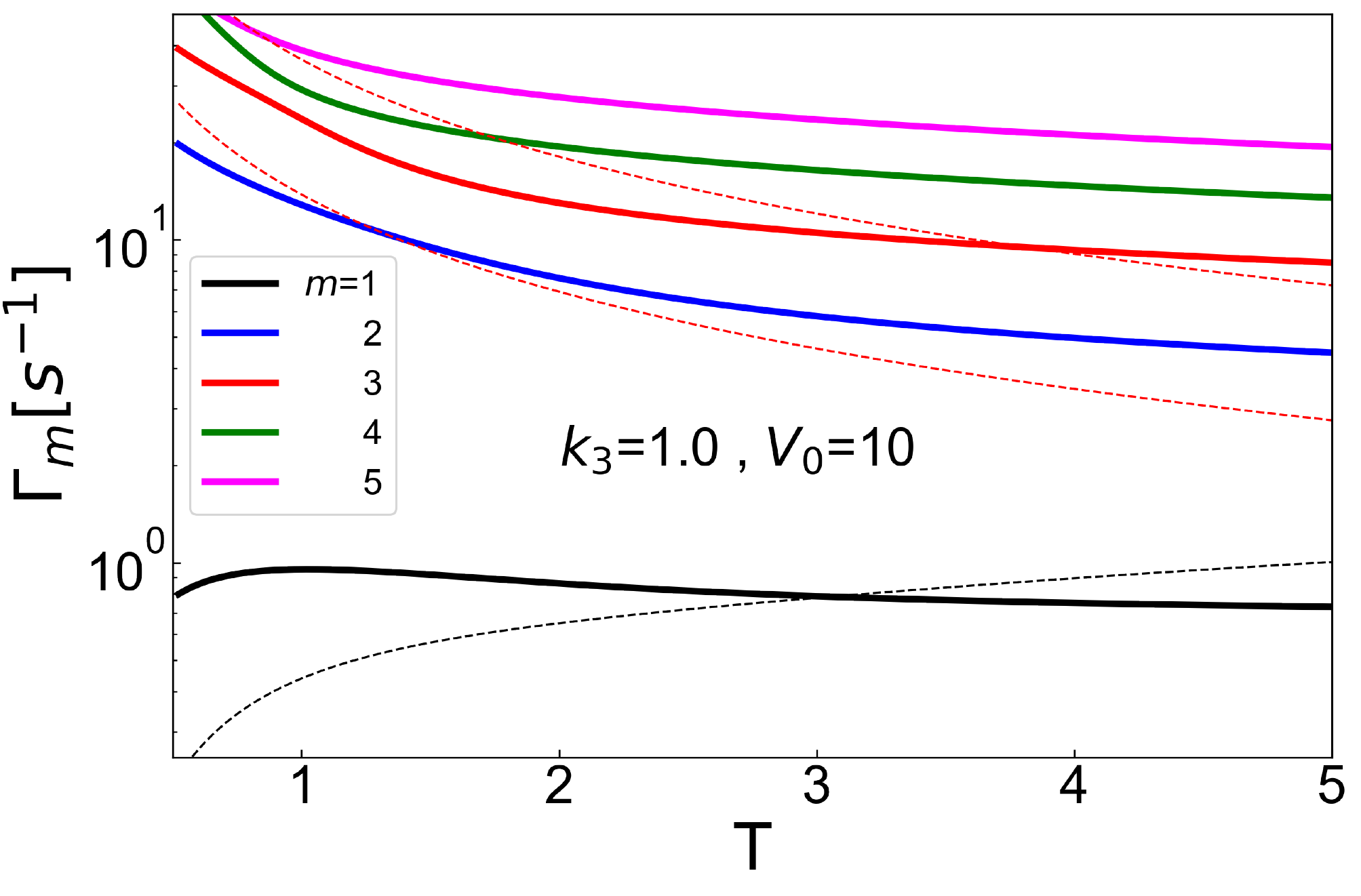}
\vspace{-0.25cm}
\caption{Smallest rates $\G_m$, obtained as eigenvalues of the FP-operator. The dashed black line is the Kramers rate for the same parameters and the red dashed lines are the rates according to the Ornstein-Uhlenbeck model for the intra-well relaxation\cite{vkamp81}.
}
\label{Figure7}
\end{figure}

In Fig.\ref{Figure6}c), the characteristic temperatures for the third-order susceptibility are shown for various values of $k_3$.
For small $k_3$ the results do not depend on the potential strength but with increasing $k_3$ the deviations from the two-state model become increasingly larger.
These findings are in accord to what can be observed in Fig.\ref{Figure6}b).
\\
{\bf 2. Dynamics}\\
We now discuss some dynamical properties of the model, in particular for the case of not too high barriers.
As mentioned above, the smallest rate (the largest eigenvalue) is associated with the barrier crossing kinetics and for large barriers coincides with the transition rate of the two-state model.
The remaining rates are those for the intra-well relaxation and for small barrier heights the time scale separation ist not perfect.
This is exemplified in Fig.\ref{Figure7}, where we show the smallest rates as a function of temperature for 
$k_3=1.0$, $V_0(T=1)=10$, yielding a barrier height of $V\,/\,T\simeq5.4$ and an asymmetry $\D\simeq9.3$.
It is evident that the Kramers approximation does not work and that there is no clearcut time scale separation in this case.

\begin{figure}[h!]
\begin{minipage}{0.475\textwidth}
\centering
\vspace{-7.0cm}
\subfigure[$\chi_1''(\om)$ as a function of scaled frequency. 
The thin dashed line is a Lorentzian.]
{\includegraphics[width=\textwidth]{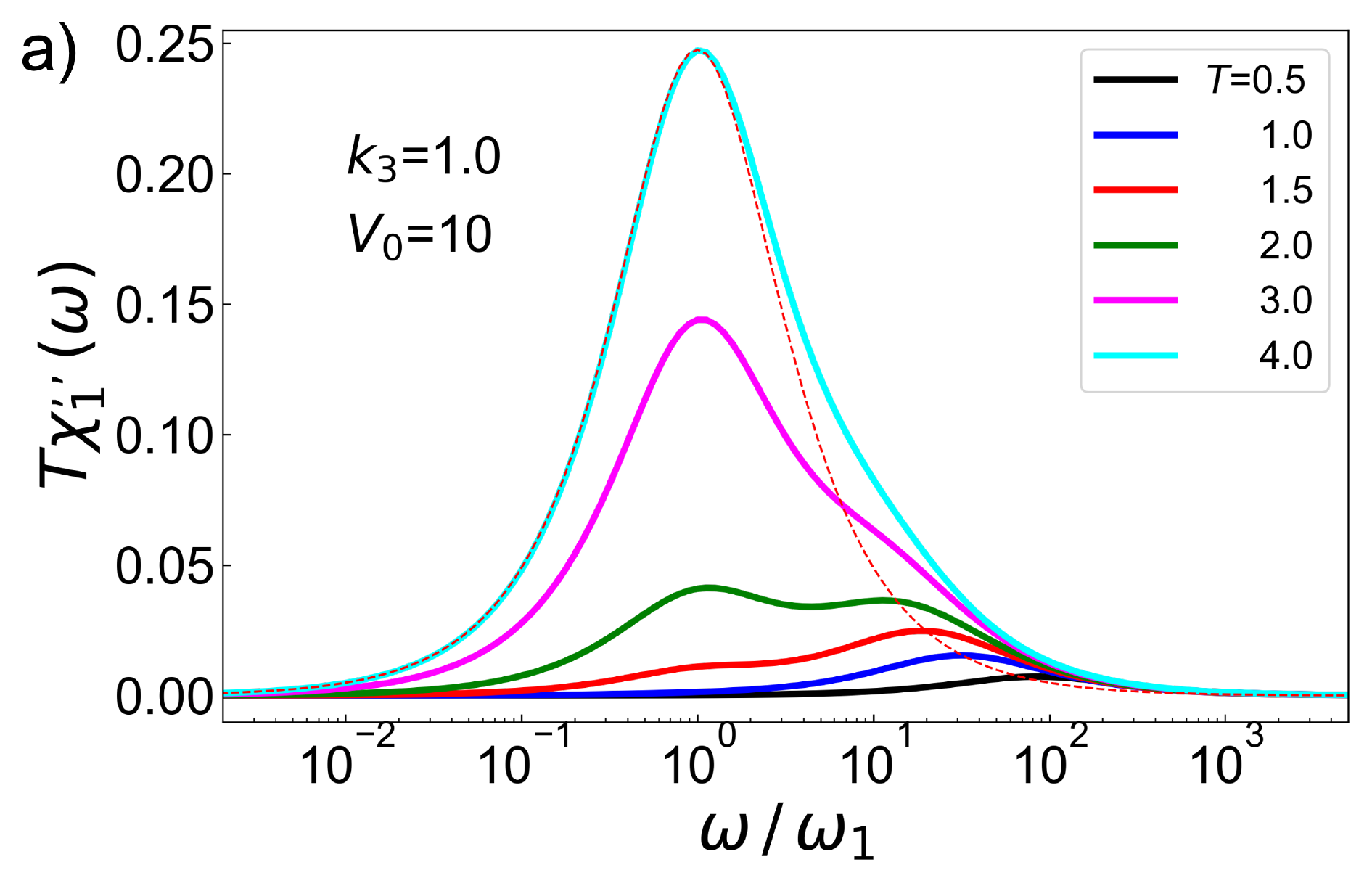}}
\end{minipage}
\begin{minipage}{0.5\textwidth}
\subfigure[$|\hat\chi_n(\om)|=|\chi_n^{n}(\om)|/|\chi_n^{n}(0)|$ as a function of scaled frequency.
Dashed lines represent the results for the two-state model: for $|\hat\chi_1(\om)|$ it is independent of temperature,
for $|\hat\chi_3^{3}(\om)|$, we used $T=2$ and for $|\hat\chi_5^{5}(\om)|$ all temperatures are presented.]
{\includegraphics[width=\textwidth]{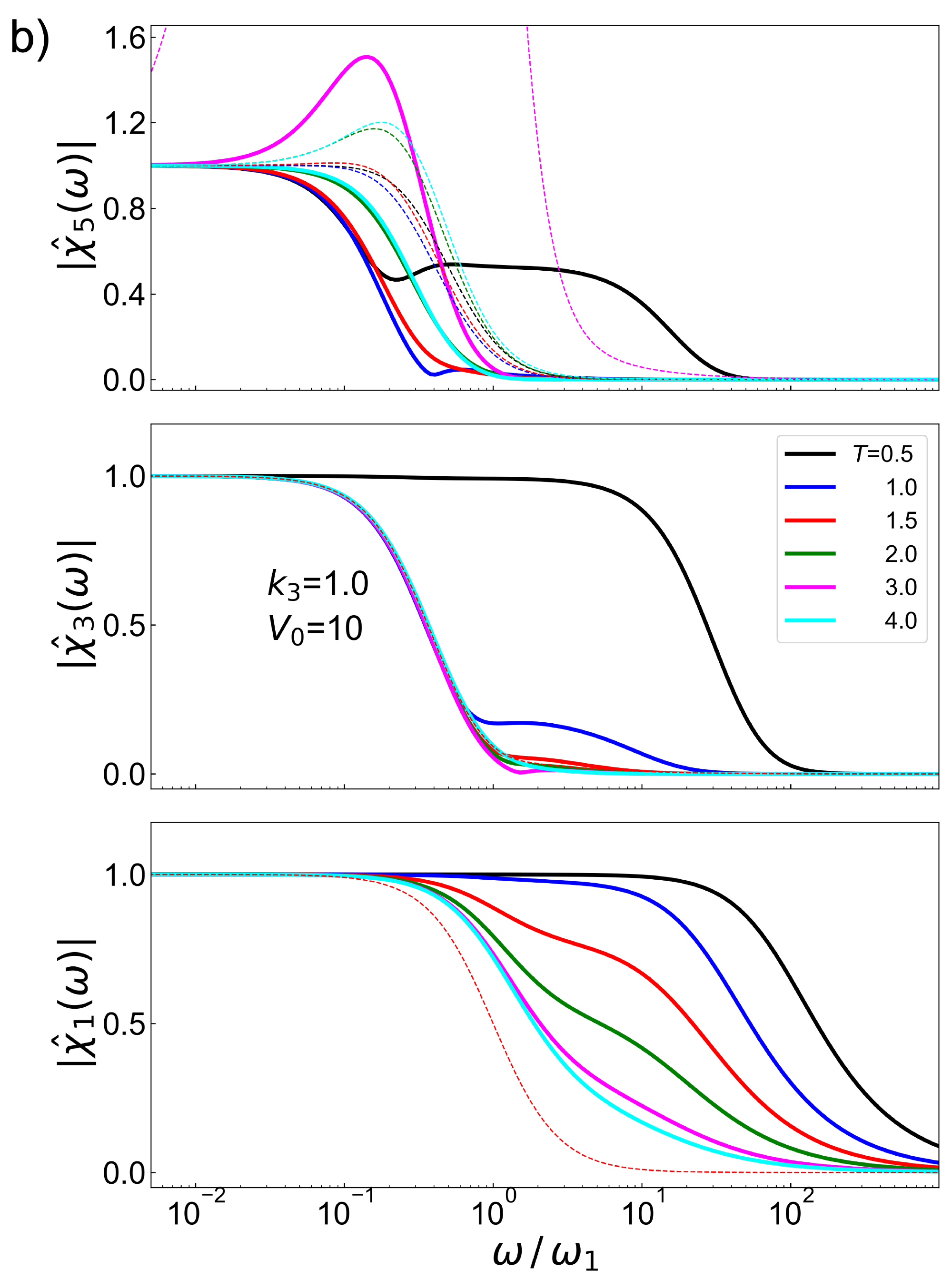}}\\
\end{minipage}
\begin{minipage}{0.475\textwidth}
\vspace{-7.0cm}
\subfigure[$\om_n$ as defined in eq.(\ref{om.n.def}) for temperatures well below the respectiv characteristic temperatures.]
{\includegraphics[width=\textwidth]{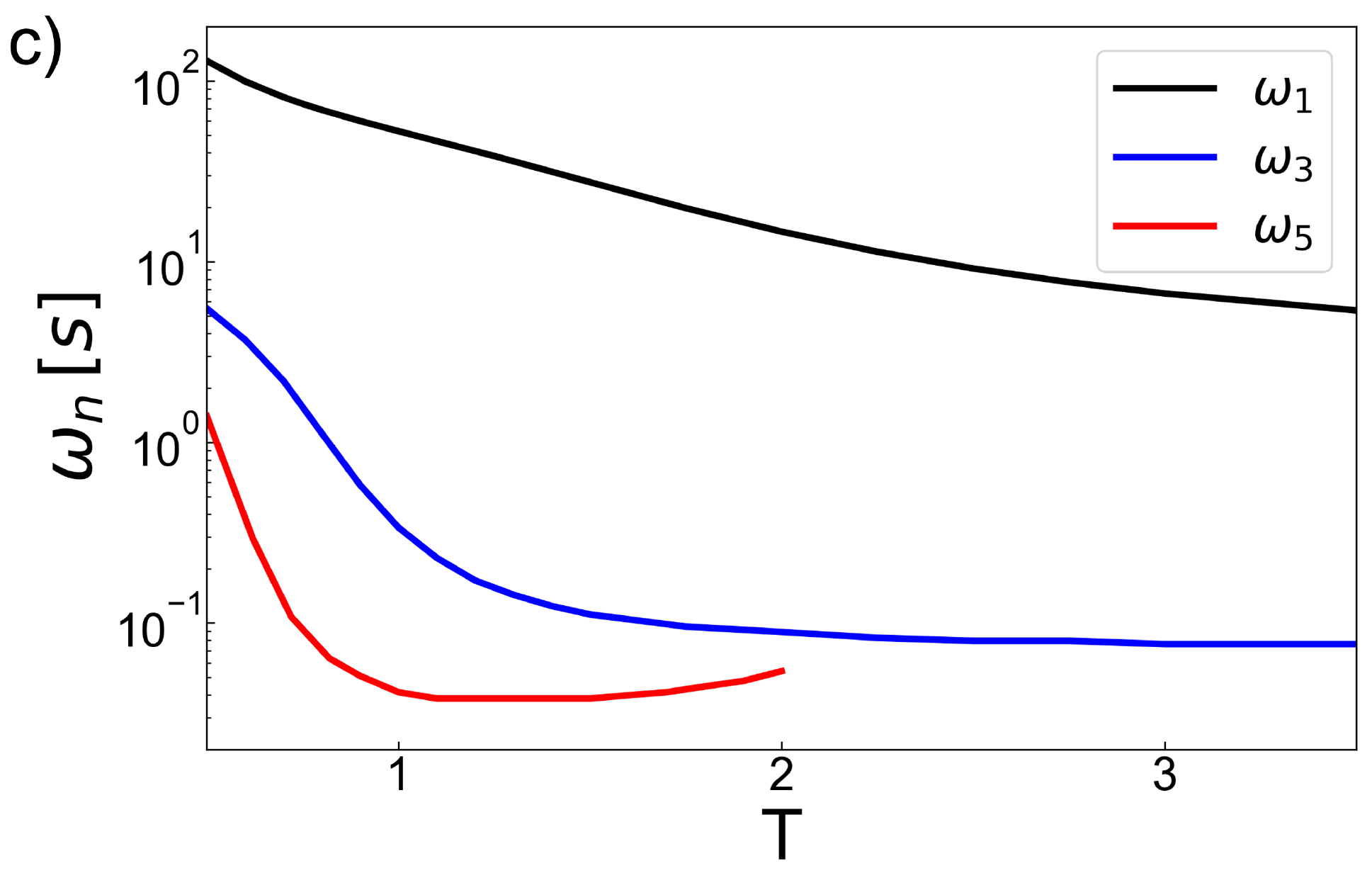}}
\end{minipage}
\caption{Imaginary part of the linear susceptibility and scaled moduli as a function of scaled frequency and 
effective relaxation frequencies as a function of temperature.
}
\label{Figure8}
\end{figure}

In Fig.\ref{Figure8}a), we present the imaginary part of the linear susceptibility, $\chi_1''(\om)$, as a function of frequency for different temperatures for the same parameters.
For low temperatures, the spectrum has a maximum at a frequency much higher than the inverse barrier crossing time.
This is because the intra-well transitions are much faster and therefore have a larger impact at low temperatures.
With increasing temperature the process of barrier crossing becomes more important and the spectrum exhibits a double-peak structure.
At high temperatures, the two-state model becomes applicable (dashed line in Fig.\ref{Figure8}a)) because the intra-well relaxation only plays a minor role.
This behavior is reminiscent of the dielectric spectra of glassforming liquids exhibiting a secondary relaxation.
In some models for the dynamics of such systems, it is assumed that the primary relaxation is governed by transitions among the minima of the free energy landscape and the secondary relaxation is related to intra-well transitions\cite{G28}.

It is interesting to consider also the nonlinear response for the same set of parameters in order to investigate if higher-order susceptibilities behave similar to the linear one.
In Fig.\ref{Figure8}b) we present the scaled moduli of $\chi_n^{n}(\om)$ for $n=1$, $3$, $5$.
It is obvious that the different relaxation processes occur with different weight in the various orders of the response.
Only at the lowest temperatures, the intra-well relaxation contributes significantly to the third-order and the fifth-roder susceptibility.
This is different for the linear response, where this process plays a significant role also at higher temperatures.
For $|\hat\chi_3(\om)|$ all temperatures are below the characteristic temperature $T_3\simeq7.06$ and for $|\hat\chi_5(\om)|$
the hump associated with $T_{5;1}\simeq2.96$ is clearly observable ($T_{5;2}\simeq11.04$ is much higher).

As a very rough estimate of the time scales relevant we use the definition of an integral over the normalized response,
\be\label{om.n.def}
\om_n={1\over\pi}\int_{-\infty}^\infty\!d\om|\hat\chi_n(\om)|
\ee
and show the results in Fig.\ref{Figure8}c).
In the vicinity of a characteristic temperature, $|\hat\chi_n(\om)|$ begins to develop a peak-like structure and the definition 
(\ref{om.n.def}) ceases to be useful.
It is, however, obvious that in the nonlinear susceptibilities the barrier crossing process has more "spectral weight" than the intra-well relaxation processes.
This is different for the linear response, where a gradual crossover from the faster intra-well relaxation to the inter-well transitions is observed.
Thus, this simple example shows that combining the measurement of linear and nonlinear response functions might allow to probe
different time scales of relaxation processes.
\section*{V. Conclusions}
We have studied the dielectric response of dipoles reorienting in a one-dimensional ADWP.
The static response in the linear regime coincides with the Langevin function if the symmetric model is considered.
Nonlinear contributions differ also in this case from the predictions of equilibrium statistical mechanics.
More important in the general case, however, is the different behavior of the nonlinear contributions to the susceptibility in case of a finite asymmetry. 
We find that in each order $n$ in the external field, the $n^{\rm th}$-order susceptibility $\D\chi_n$ changes its sign at $(n-1)/2$ characteristic temperatures $T_{n\a}$, an unexpected result for non-interacting dipoles.
Therefore, we conclude that the model has to be applied with care for the interpretation of results obtained from nonlinear dielectric spectroscopy.

In order to assure that these findings are not artefacts of the two-state approximation usually employed when considering the ADWP model, we solved the FPE for the Brownian motion in a model potential consisting of a harmonic, a cubic and a quartic term.
For a linear dependence of the dipole moment on the coordinate, the results coincide with those of the two-state model in the limit of high barriers, as expected on general grounds.
For small barriers, we find a very similar behavior of the static response functions with the characteristic temperatures shifted to lower temperature.
Only the observed decrease of the characteristic temperature with increasing asymmetry is at variance with the two-state model, where an increase is predicted.
The overall features are, however, not altered significantly.

We also considered the frequency-dependent susceptibilities up to fifth order in the external field for an ADWP with a small barrier, such that the time scale separation between the intra-well and the inter-well transitions is not guaranteed.
We find that the linear susceptibility at low temperatures is mainly determined by the intra-well relaxation processes and only at higher temperatures the barrier crossing becomes dominant.
For the higher order susceptibilities the latter process is more important even at lower temperatures.
These findings suggest that combinations of linear and nonlinear susceptibilities might be employed in order to resolve various relaxation mechanisms 
such as the inter-well and intra-well transitions in the ADWP model.
One might anticipate that a somwhat similar behavior can also occur for models with more complex energy landscapes.
It would be interesting to study further models exhibiting more than a single time scale and compare the results for the various nonlinear response functions.
\section*{Acknowledgement}
I thank Roland B\"ohmer, Gerald Hinze, and Jeppe Dyre for fruitful discussions.
\begin{appendix}
\section*{Appendix A: Characteristic temperatures in the two-state model}
\setcounter{equation}{0}
\renewcommand{\theequation}{A.\arabic{equation}}
In this Appendix, we determine the characteristic temperatures $T_{n,\a}$ starting from the Taylor-expansion of the argument of
eq.(\ref{Chi.stat}),
\be\label{Meq.delta}
P=\sum_{n\, \rm uneven}\!\!H^n\D\chi_n
\quad\mbox{with}\quad
\D\chi_n=\b^n\lg M^{n+1}\rg\left(1-\d^2\right)\Pi_n(\d)
\ee
where $\lg M^k\rg=M^k\lg\cos^k{(\theta)}\rg$ ($=M^k/(k+1)$ for $k$ even in 3d).
The $\Pi_n(\d)$ are polynomials in $\d^2$ of degree $(n-1)$ and the first few are given by:
\Be\label{Dchi.k}
&&\Pi_1=1\nonumber\\
&&\Pi_3={1\over3}(-1+3\d^2)\nonumber\\
&&\Pi_5={1\over15}(2-15\d^2+15\d^4)\\
&&\Pi_7={1\over315}(-17+231\d^2-525\d^4+315\d^6)\nonumber\\
&&\Pi_9={1\over2835}(62-1320\d^2+5040\d^4-6615\d^6+2835\d^8)
\nonumber
\Ee
The solutions of $\Pi_n(\d)=0$ yield $(n-1)/2$ characteristic temperatures $T_{n,\a}$ for which 
$\D\chi_n(T_{n,\a})=0$.
These temperatures are directly related to the positive roots $\d_{n,\a}$ via
\be\label{Tna.def}
T_{n,\a}={\D\over\ln{\left\{(1+\d_{n,\a})/(1-\d_{n,\a})\right\}}}
\ee
The characteristic temperatures for the third order and the fifth order susceptibilities are given in eq.(\ref{T.3.5}).
For the next ones, one finds:
\Be\label{T.7.9}
&&T_{7;1}\simeq0.2130\D \,\,;\,\, T_{7;2}\simeq0.4764\D\,\,;\,\, T_{7;3}\simeq1.6001\D\nonumber\\
&&T_{9;1}\simeq0.161\D \,\,;\,\, T_{9;2}\simeq0.318\D \,\,;\,\, T_{9;3}\simeq0.625\D \,\,;\,\, T_{9;4}\simeq2.008\D
\Ee
\section*{Appendix B: Higher-order susceptibilities}
\setcounter{equation}{0}
\renewcommand{\theequation}{B.\arabic{equation}}
As it has been discussed e.g. in ref.\cite{Albert:2019}, the higher-order susceptibilities fulfill certain symmetry relations yielding combinatorical prefactors for the various frequency components.
We use the same expansion as in eq.(\ref{Meq.delta}) also for the frequency-dependent susceptibilities,
\be\label{Pn.om}
P(\om)=\sum_{n}\!P_n(\om)=\sum_{n}\!H^n\chi_n(\om)
\ee
with uneven $n$.
For the much discussed third-order and fifth-order, one has, cf.\cite{Albert:2019,Albert:2016}
\Be\label{Chi.3.5}
&&\chi_3(\om)={1\over4}\left(3\chi_3^{(1)}(\om)+\chi_3^{(3)}(\om)\right)\nonumber\\
&&\chi_5(\om)={1\over16}\left(10\chi_5^{(1)}(\om)+5\chi_5^{(3)}(\om)+\chi_5^{(5)}(\om)\right)
\Ee
Here, $\chi_n^{(k)}(\om)$ denotes the $k\om$-component of the $n$th-order susceptibility.
Similarly, one finds:
\Be\label{Chi.7.9}
&&\chi_7(\om)={1\over64}\left(35\chi_7^{(1)}(\om)+21\chi_7^{(3)}(\om)+7\chi_7^{(5)}(\om)+\chi_7^{(7)}(\om)\right)\nonumber\\
&&\chi_9(\om)={1\over256}\left(126\chi_9^{(1)}(\om)+84\chi_9^{(3)}(\om)+36\chi_9^{(5)}(\om)
+9\chi_9^{(7)}(\om)+\chi_9^{(9)}(\om)\right)
\Ee
The zero-frequency limits of the $\chi_n^{(k)}(\om)$ for a given order coincide and sum up to the same result as obtained from the
Taylor expansion of the equilibrium response $P$ according to eq.(\ref{Chi.stat}).
Note that the definitions of the various frequency components of the third-order and the fifth-order susceptibilities differ from the definitions used in refs.\cite{G75,G81,G87,G95,G92} by the corresponding combinatorical factors.
If, on the other hand, a selected $\om$-component is measured, one obtains the corresponding fraction of $\chi_n(0)$ according to 
eqns.(\ref{Chi.3.5},\ref{Chi.7.9}).
Furthermore, one has for the $k\om$-components, $P^{(k)}(\om)=\sum_{n}\!H^n\chi_n^{(k)}(\om)$,
normalized to the lowest-order contribution, cf. eqns(\ref{Chi.3.5},\ref{Chi.7.9}):
\Be\label{Chi.comp}
&&P^{(1)}(\om)=\hat\chi_1^{(1)}(\om)+{3\over4}\hat\chi_3^{(1)}(\om)+{5\over8}\hat\chi_5^{(1)}(\om)
+{35\over64}\hat\chi_7^{(1)}(\om)+{63\over128}\hat\chi_9^{(1)}(\om)+\cdots\nonumber\\
&&P^{(3)}(\om)=\hat\chi_3^{(3)}(\om)+{5\over4}\hat\chi_5^{(3)}(\om)+{21\over16}\hat\chi_7^{(3)}(\om)+{21\over16}\hat\chi_9^{(3)}(\om)+\cdots\\
&&P^{(5)}(\om)=\hat\chi_5^{(5)}(\om)+{7\over4}\hat\chi_7^{(5)}(\om)+{9\over4}\hat\chi_9^{(5)}(\om)+\cdots\nonumber\\
&&\cdots\nonumber
\Ee
where we abbreviated $\hat\chi_n^{(k)}(\om)=H^n\chi_n^{(k)}(\om)$.
\section*{Appendix C: Properties of a specific ADWP model}
\setcounter{equation}{0}
\renewcommand{\theequation}{C.\arabic{equation}}
{\bf General definitions:}\\
Using the abbreviations $\bar V(q)=V_{ADWP}(q)/(V_0k_4)$ and $\k_l=(k_l/k_4)$, $l=2,3$ one has for $V_{ADWP}(q)$ according to 
eq.(\ref{V.ADWP}):
\be\label{V.bar}
\bar V(q)={1\over4}q^4-{\k_2\over2}q^2+{\k_3\over3}q^3,
\ee
cf. eq.(\ref{V.ADWP}). 
We will use only positive constants $k_l$.
In this case, the minima of the potential are located at
\be\label{q.1.2}
q_{1/2}=-{1\over2}\left(\k_3\pm w\right)
\quad\mbox{with}\quad
w:=\sqrt{4\k_2+\k_3^2}
\ee
and the maximum is at $q_T=0$.
For the computation of the activation energies needed in the expressions for the Kramers rates, the following definitions will be used, cf. Fig.(\ref{Fig1}):
\be\label{V1.V2.def}
V_{1/2}=V_{ADWP}(q_{1/2})=E_0\mp\D/2\,,\, V_T=V_{ADWP}(q_T)=0
\ee
yielding the 'ground state energy', the asymmetry and the barrier:
\be\label{V.1.2}
\bar E_0=-{1\over24}\left\{ 6\k_2^2+6\k_2\k_3^2+\k_3^4\right\}
\,\,,\,\,
\bar\D={1\over12}\k_3w^3
\,\,,\,\,
\bar V=\bar V_T-{1\over2}\left(\bar V_1+\bar V_2\right)=-\bar E_0
\ee
In addition, the curvatures at the extrema are given by:
\be\label{krumm}
\bar V''(q_{1/2})={1\over2}\left\{4\k_2+\k_3^2 \pm \k_3w\right\}
\,,\,
\bar V''(q_T)=-\k_2.
\ee
{\bf Kramers rates and two-state approximation:}\\
The inverse mean first passage times for reaching the barrier located at $q_T$ from either $q_1$ or $q_2$,
$W_{T,1}$ and $W_{T,2}$, according to Kramers theory are given by:
\be\label{W.Tl.Kramer}
W_{T,1}=W_{T,1}^{0}e^{-\b\D/2} \,,\, W_{T,2}=W_{T,2}^{0}e^{+\b\D/2}
\quad\mbox{with}\quad
W_{T,l}^0=D{\sqrt{|V''(q_T)|V''(q_l)} \over 2\pi}e^{-\b V}
\ee
where $D$ denotes the diffusion coefficient.
The transition rates for the inter-well transitions are related to these (using a steady state approximation for the transition state) via\cite{Schulten:1981,G67}:
\be\label{W.kl.res}
W_{21}^{-1}=W_{T,1}^{-1}+W_{T,2}^{-1}{Z_1\over Z_2}
\quad; \quad
W_{12}^{-1}=W_{T,2}^{-1}+W_{T,1}^{-1}{Z_2\over Z_1}
\ee
where $Z_k=\int_{q\in k}\!\!dq\,e^{-\b V_{ADWP}(q)}$ is the partition sum restricted to well $k=1,2$.
Additionally,, the populations of the two 'states' $1$ and $2$ are given by the expression:
\be\label{pk.t.def}
p_k(t)=\int_{q\in k}\!\!\!dq\,p(q,t)
\ee
This yields the ME, ${\dot p}_k(t)=-W_{lk}p_k(t)+W_{kl}p_l(t)$, for $l,k=1,2$, cf. ref.\cite{Schulten:1981}.

In the harmonic approximation, $V_{ADWP}(q)\simeq V_k+{1\over2}V''(q_k)(q-q_k)^2$ near the minimum located at $q_k$, one finds:
\be\label{Zk.harm}
Z_k\simeq {\sqrt{2\pi}\over\sqrt{\b V''(q_k)}}e^{-\b V_k}
\ee
In this approximation, one finds for the long-time limit of the populations of the states, $p_k^{\rm eq}=Z_k/Z$, cf. eq.(\ref{pk.t.def}).
For the transition rates, eqns.(\ref{W.Tl.Kramer},\ref{W.kl.res}) give:
\be\label{W.kl.harm.R}
W_{kl}={1\over2}W_{T,l}
\quad\mbox{and}\quad
R_{\rm Kramers}=W_{12}+W_{21}={1\over2}\left(W_{T,1}+W_{T,2}\right)
\ee
cf. eq.(\ref{R.H0}).
This means, that in the present approximation the transition rate from one well to the other is just given by half the inverse of the mean first passage time to the barrier.
This is meaningful, because after reaching the barrier one has exactly the same probability for leaving it to one of the two wells.
It is obvious that in the two-state model discussed in the text, the dependence of $W_{kl}^0$ on the curvature in the initial state of the transition has been neglected.
If one approximates the curvatures in the minima by their mean,
\[
V''_m={1\over2}\left(V''(q_1)+V''(q_1)\right)={1\over2}\left\{4\k_2+\k_3^2 \right\}\cdot V_0
\]
one finds $R=W_{12}+W_{21}=2W\cosh{\!(\b\D/2)}$ with
\be\label{W.vgl}
W=W_0e^{-\b V}=D{\sqrt{|V''(q_T)|V''_m} \over 4\pi}e^{-\b V}
\ee
The error of the approximations is negligible for $k_3\ll k_2$.

{\bf Static nonlinear susceptibilities:}\\
The static polarization is given by the expectation value of the moment in a finite external a.c. field:
\be\label{DChi.FPE.gen}
P=\Bigl\lg Z_H^{-1}\int\!dqM(q) e^{-\b(V(q)-M(q)\cdot H)}\Bigl\rg
\ee
where $V(q)$ denotes the ADWP, $Z_H=\int\!dq e^{-\b(V(q)-M(q)\cdot H)}$ and $\lg\cdot\rg$ is meant to include an angular average.
The expansion of $P$ in powers of the field $H$ is given by
\be\label{DChi.FPE.expand}
P={\sum_n{\b^n\over n!}M^{(n+1)}H^n\over\sum_n{\b^n\over n!}M^{(n)}H^n}
\quad\mbox{with}\quad
M^{(n)}=\int\!dq M(q)^n e^{-\b V(q)}
\ee
As noted in the text, in the actual calculations we employed the linear relationship given in eq.(\ref{M.lin.q}).
Using the fact that $M^{(0)}=Z$ and the definition of the averages, $\lg M^n\rg =M^{(n)}/Z$, one finds for the susceptibilities:
\be\label{DChi.n.FPE}
\D\chi_n={\b^n\over n!}\k_n(M)
\ee
where $\k_n(M)$ denotes the $n$th order cumulant of the distribution of $M$-values\cite{Matyushov:2015}.

We note that the approximations used to derive the Kramers rates, the two-state model and the moments are independent of the particular form of the ADWP employed in the present calculations.
\end{appendix}
\end{document}